\documentclass[cernpreprint,texlive=2011,texmf,txfonts,USenglish]{atlasdoc} 
\pdfoutput=1
\usepackage[subfigure=true,biblatex=false]{atlaspackage} 
\usepackage{atlasphysics} 
\usepackage{lipsum}
\usepackage{epstopdf}
\usepackage{cite}
\newcommand{\papertitle}{Measurement of the top pair production cross-section in 8~TeV proton--proton collisions using kinematic information in the lepton+jets final state with ATLAS}

\AtlasTitle{\papertitle}
\author{The ATLAS Collaboration}
\AtlasJournal{Phys. Rev. D.}
\PreprintIdNumber{CERN-PH-EP-2015-024}
\AtlasAbstract{
A measurement is presented of the $t\bar{t}$ inclusive production cross-section
in $pp$ collisions at a center-of-mass energy of $\sqrt{s}=8$~TeV using data collected
by the ATLAS detector at the CERN Large Hadron Collider. The
measurement was performed in the lepton+jets final state using a data set
corresponding to an integrated luminosity of 20.3~fb$^{-1}$.  The cross-section
was obtained using a likelihood discriminant fit and $b$-jet
identification was used to improve the signal-to-background ratio. The
inclusive $t\bar{t}$ production cross-section 
was measured
to be $260\pm 1{\mathrm{(stat.)}} ^{+22}_{-23} {\mathrm{(syst.)}}\pm 8{\mathrm{(lumi.)}}\pm 4{\mathrm{(beam)}}$~pb
assuming a top-quark mass of
172.5~GeV, in good agreement with the theoretical prediction of
$253^{+13}_{-15}$~pb.  The $\ttbar \to (e,\mu)+{\mathrm{jets}}$ production
cross-section in the fiducial region determined by the detector acceptance is
also reported.
}

\begin{document}

\maketitle

\section{\label{sec:intro}
Introduction
}
The measurement of the top-quark pair production cross-section is an important
part of the physics program of the Large Hadron Collider
(LHC)~\cite{lhc_machine}. As the top-quark is the heaviest known fermion, it
has the largest coupling to the recently discovered Higgs boson and plays a
special role in many theories beyond the Standard Model (SM).
New physics may result in additional top-quark decay channels or
new mechanisms of $t\bar{t}$ production that can cause
the measured cross-section to deviate from the SM prediction.
Also, $t\bar{t}$ production is the dominant background to
many searches for new physics.

The inclusive $pp\to t\bar{t}$ cross-section $\sigma_{t\bar{t}}$ at a
center-of-mass energy of $\sqrt{s}=8$~TeV has been calculated to be
$253^{+13}_{-15}$~pb for a top-quark mass of
$m_{\rm top}=172.5$~GeV. 
These calculations were performed
at next-to-next-to-leading order (NNLO) in QCD including resummation of
next-to-next-to-leading logarithmic (NNLL) soft gluon terms using the program
top++~2.0~\cite{TTcalc01,TTcalc02,TTcalc03,TTcalc04,TTcalc05,TTcalc06}.  In
these calculations, the renormalization scale $\mu_{\rm R}$ and factorization
scale $\mu_{\rm F}$ were both set equal to $m_{\rm top}$.
The NNLO+NNLL value was found to be about 3\% larger than the NNLO-only
calculation as implemented in Hathor~1.5~\cite{TopHATHOR2011}.
As a result of recent progress in theoretical calculations, the $t\bar{t}$
production cross-section has been evaluated with uncertainties below 6\%. In
calculating these uncertainties, the scale uncertainty, evaluated by
independent variations of $\mu_{\rm R}$ and $\mu_{\rm F}$ by factors of $1/2$
and $2$, was combined in quadrature with uncertainties on the coupling strength
of the strong interaction, $\alpha_{\rm S}$, and the parton distribution functions
(PDFs). The latter were calculated using the PDF4LHC
prescription~\cite{TTPDF2011} with MSTW2008 NNLO (at the 68\%
confidence level)~\cite{TTPDF2009, TTPDFAs2009},
CT10 NNLO~\cite{TTPDF2010, TTCT10NNLO2013}
and NNPDF2.3 5f FFN~\cite{LHCPDF2013} PDF sets.

In the SM, the top-quark decays into a $W$-boson and a $b$-quark with a
branching ratio close to 100\%. The present analysis aims to measure the
$t\bar{t}$ production cross-section in the lepton+jets final state, where one
of the $W$-bosons decays into an electron or a muon (collectively called a
lepton and denoted $\ell$) and a corresponding neutrino (including the
$W\to\tau\nu_\tau\to e\nu_\tau\nu_e$ and
$W\to\tau\nu_\tau\to\mu\nu_\tau\nu_\mu$ decays), and the other $W$-boson decays
hadronically. The final state is characterized by the presence of a highly
energetic isolated lepton, large missing transverse momentum ($\MET$) due to the
neutrino(s) escaping detection, and four jets due to the two $b$-quarks from the
top-quark decays and the two quarks from the hadronic $W$ decay. The selected
events are required to have at least three jets, allowing one jet to be
undetected due to limited detector acceptance and jet reconstruction
inefficiency.  At least one of the jets is further required to be identified
as a $b$-jet. 

The inclusive top-quark pair production cross-section in $pp$ collisions at
8~TeV has been previously measured in the dilepton channel by both the ATLAS and
CMS Collaborations~\cite{ATLAS_8TeV, CMS_8TeV}. The results are found to be in good agreement
with theoretical predictions.
This paper presents the measurement of the $\ttbar$ production
cross-section of the ATLAS Collaboration in the lepton+jets channel,
which provides a cross-check of the dilepton measurement
that has different background conditions and systematic uncertainties,
and contributes to the combined $\ttbar$ production cross-section result.
In addition, the
measurement in the lepton+jets channel is important for probing
the presence of new physics that changes the top-quark decay branching
fractions, e.g.
production of charged Higgs bosons $H^+$ via $t\to H^+ b$ decays.

The measurement was performed 
using the full 8~TeV data set (20.3~\ifb).
In addition to the total $\ttbar$ production
cross-section, a fiducial cross-section was measured, defined using physics
objects constructed of stable particles to approximate the
$\ttbar\to\ell\nu$+jets detector acceptance.

The majority of the events in the selected sample originate from $t\bar{t}$
production. However, events from other SM processes ($W/Z$+jets,
single top-quark, diboson, and multijet production) are also present.
To determine the signal fraction, a discriminating variable
(likelihood discriminant, LHD) was constructed based on the kinematic
properties of Monte Carlo (MC) simulated signal and background events.  A
weighted sum of LHD distributions (``templates'') for the signal and for the
background was fitted to the LHD distribution in data, and the resulting number
of signal events was converted to the $\ttbar$ inclusive production
cross-section using the signal-reconstruction efficiency (determined from MC simulation),
known branching ratio for the lepton+jets final state, and the total integrated
luminosity.

The paper is organized as follows. The ATLAS detector is briefly described in
Sec.~\ref{sec:detector}, followed by descriptions of the data and MC samples
used in the analysis (Sec.~\ref{sec:data}). The event selection and the
backgrounds are outlined in Sec.~\ref{sec:selection}. The measurement
procedure is presented in Sec.~\ref{sec:analysis} for
the inclusive case, and in Sec.~\ref{sec:fiducial} for the fiducial case.
The estimation of the
systematic uncertainties is presented in Sec.~\ref{sec:systematics}.
Finally, the results and conclusions are given in Sec.~\ref{sec:conclusions}.

\section{\label{sec:detector}
The ATLAS detector
}
The ATLAS detector~\cite{atlasdetector} at the LHC covers nearly the entire solid angle around the collision point.
It consists of an inner tracking detector surrounded by a thin superconducting solenoid, electromagnetic and hadronic calorimeters,
and a muon spectrometer incorporating three large superconducting toroid magnets.
The inner-detector system (ID) is immersed in a \SI{2}{\tesla} axial magnetic field 
and provides charged particle tracking in the range $|\eta| < 2.5$.

The high-granularity silicon pixel detector 
is closest to the interaction region
and typically provides three measurements per track, 
the first hit being normally in the innermost layer.
It is followed by the silicon microstrip tracker 
which has four layers in the barrel region.
These silicon detectors are complemented by the transition radiation tracker,
which enables radially extended track reconstruction up to $|\eta| = 2.0$. 
The transition radiation tracker also provides electron identification information 
based on the fraction of hits (typically 30 in total) above a higher energy deposit threshold corresponding to transition radiation.

The calorimeter system covers the pseudorapidity range $|\eta| < 4.9$.
Within the region $|\eta|< 3.2$, electromagnetic calorimetry is provided by barrel and 
endcap high-granularity lead/liquid-argon (LAr) electromagnetic calorimeters,
with an additional thin LAr presampler covering $|\eta| < 1.8$,
to correct for energy loss in material upstream of the calorimeters.
Hadronic calorimetry is provided by the steel/scintillating-tile calorimeter,
segmented into three barrel structures within $|\eta| < 1.7$, and two copper/LAr hadronic endcap calorimeters.
The solid angle coverage is completed with forward copper/LAr and tungsten/LAr calorimeter modules
optimised for electromagnetic and hadronic measurements respectively.

The muon spectrometer (MS) comprises separate trigger and
high-precision tracking chambers measuring the deflection of muons in a magnetic field generated by superconducting air-core toroids.
The precision chamber system covers the region $|\eta| < 2.7$ with three layers of monitored drift tubes,
complemented by cathode strip chambers in the forward region, where the background is highest.
The muon trigger system covers the range $|\eta| < 2.4$ with resistive plate chambers in the barrel, and thin gap chambers in the endcap regions.
A three-level trigger system is used to select interesting events~\cite{PERF-2011-02}.
The Level-1 trigger is implemented in hardware and uses a subset of detector information
to reduce the event rate to a design value of at most \SI{75}{\kHz}.
This is followed by two software-based trigger levels which together reduce the event rate to about \SI{200}{\Hz}.

\section{\label{sec:data}
Data and simulated samples
}
The data were collected during the 2012 LHC running period
at a center-of-mass energy of 8$\TeV$.
After applying data-quality selection criteria, the data set used
in the analysis corresponds to an integrated luminosity of 20.3~\ifb.

Simulated $\ttbar$ events were generated with {\sc Powheg}~\cite{FRI-0701}
interfaced to {\sc Pythia} v6.426~\cite{PYT6} for the fragmentation
and hadronization, with the next-to-leading order (NLO) CT10 PDF set. 
The Perugia2011C underlying event
tune~\cite{Perugia2011C} with the CTEQ6L1 PDF set~\cite{CTEQ2008} was used.  An alternative
$\ttbar$ MC sample was produced with {\sc MC@NLO} v4.01~\cite{FRI-0201} using
the CT10 PDF set. In this sample, the parton shower and the underlying event
simulations were performed with {\sc Herwig} v6.520~\cite{COR-0001} and {\sc
Jimmy} v4.31~\cite{Butterworth:1996zw}, respectively, using the AUET2
tune~\cite{PUB-2011-008}. To estimate the model dependence of the parton shower
and fragmentation modeling, the signal sample generated with {\sc
Powheg} interfaced to {\sc Pythia} was compared to that generated with {\sc Powheg} interfaced
to {\sc Herwig+Jimmy}.  Additional $\ttbar$ samples simulated with {\sc
AcerMC}~\cite{AcerMC} interfaced to {\sc Pythia} based on the CTEQ6L1 PDF set were
used to study the systematic uncertainties arising from initial- and final-state
radiation. All $\ttbar$ samples were produced with a top-quark mass of
172.5~GeV and normalized to the NNLO+NNLL cross-section quoted in
Sec.~\ref{sec:intro}.

The dominant background to $\ttbar$ production is vector-boson
production in association with jets, $V$+jets ($V=W,Z/\gamma^*$).
Samples of events were generated
using {\sc Alpgen}~\cite{ALPGEN} interfaced to {\sc Pythia} v6.426
based on the CTEQ6L1 PDF set and the Perugia2011C tune. The
MLM parton and jet matching procedure~\cite{alpgen_MLM} was applied inclusively for 
$V$+5-light-partons (2$\rightarrow$7) production and exclusively for 
lower multiplicity samples. In addition to $V$+light partons, the
production of vector-bosons with additional heavy-flavor partons 
($V+c$+jets, $V+c\bar{c}$+jets, $V+b\bar{b}$+jets) was also simulated. 
Inclusive $V$+jets
samples were formed by combining the samples of light and heavy quarks according to
their respective cross-sections.  An alternative $W$+jets sample simulated with
{\sc Sherpa}~\cite{Sherpa} with massive $b$/$c$-quarks was produced to
evaluate systematic uncertainties on $W$+jets modeling.
The $V$+jets processes were
normalized to the NNLO calculations~\cite{Hamberg:1990np,Anastasiou:2003ds}.

The background contribution from single-top production, including $t$- and
$s$-channel contributions and $Wt$ production was simulated with {\sc Powheg}
interfaced to {\sc Pythia} using the CT10 PDF set.  Finally, diboson production
($WW$, $WZ$, $ZZ$) was simulated with {\sc Herwig} using the CTEQ6L1 PDF set.
All samples were scaled according to their theoretical production
cross-sections: at approximate NNLO for the single top
$t$-channel~\cite{Kidonakis:2011wy}, and at NLO+NNLL for the single top
$s$-channel~\cite{Kidonakis:2010tc} and
$Wt$-production~\cite{Kidonakis:2010ux}.  
The diboson processes were normalized to the NLO~\cite{lhc_diboson} predictions.

Most of the $\ttbar$ samples and all the background samples were processed
through the full ATLAS detector simulation~\cite{AtlasSimulation}
based on GEANT4~\cite{AGO-0301}.
A few $\ttbar$ samples used for the evaluation of certain
systematic effects (modeling of initial- and final-state radiation and
parton showers -- see Sec.~\ref{sec:systematics} for details)
were produced using the ATLAS fast
simulation that employs parameterized showers in the
calorimeters~\cite{ATLAS_FastII}. All MC samples were  
reconstructed
using the same analysis chain as used for the data. Correction factors were
applied in order to better reproduce the trigger, lepton reconstruction
efficiency, and $b$-jet identification efficiency observed in the data.  The
simulations also included the effect of multiple $pp$ collisions per bunch
crossing (pileup).

\section{\label{sec:selection}
Event selection
}
The events were required to pass a logical OR of isolated and nonisolated
single-lepton ($e$ or $\mu$) trigger conditions with transverse momentum $\pT$
thresholds of 24~GeV for isolated triggers and 60~(36)~GeV for nonisolated
single-$e$ (single-$\mu$) triggers.
All events were also required to have a reconstructed hard collision primary
vertex (the main PV) built of at least five particle tracks with $\pT>0.4$~GeV.

The reconstructed objects used in the analysis include electrons, muons, jets,
and missing transverse momentum.  Electron candidates were reconstructed from
an electromagnetic energy deposit matched to a track in the inner detector~\cite{Systematics_lepton1}.  
They were required to have transverse energy
$\ET>40$~GeV and pseudorapidity of the cluster\footnote{Unlike 
the
(regular) pseudorapidity determined using the object direction, the
pseudorapidity of the cluster is defined using the position of the
reconstructed cluster in the calorimeter with respect to the geometric center
of the detector.} $|\eta|<2.47$, excluding the barrel--endcap transition region
$1.37<|\eta|<1.52$. Electron candidates were also required to originate from
less than 
2~mm along the $z$-axis from 
the main PV and satisfy  
isolation criteria. The
latter involve
a combination of 
calorimeter isolation (a requirement on a
sum of energies of calorimeter cells within a cone of size $\Delta R=0.2$
around the electron direction) and
track isolation (a requirement on a
scalar sum of track $\pT$ within a cone of size $\Delta R=0.3$, in each case excluding the contribution from the electron itself). 
Both
requirements were chosen to separately result in a 90\% electron reconstruction
efficiency for prompt electrons from $Z\to ee$ decays.  Muon candidates were
reconstructed using combined information from the muon spectrometer and the
inner tracking detectors~\cite{MuonReconstruction}. They were required to have
$\pT>40$~GeV and $|\eta|<2.5$, and, like electrons, to
originate from within 2~mm along the $z$-axis of the main PV. The muon
isolation was defined in terms of the ratio of the scalar sum of the track
$\pT$ in a cone of variable radius $\Delta R=10{\rm~GeV}/\pT^\mu$ around the
muon direction (excluding the muon track itself) to the $\pT$ of the muon
($\pT^\mu$). This ratio was required to be less than 0.05,
corresponding to a 97\% selection efficiency for prompt muons from $Z\to\mu\mu$
decays.

Jets were reconstructed using the anti-$k_t$
algorithm~\cite{antikt1,antikt2} with radius parameter $R=0.4$.  The
measured jet energy was corrected for inhomogeneities and for the
noncompensating nature of the calorimeter through jet $\pT$- and
$\eta$-dependent factors derived from MC simulation and validated with
data. Jets were calibrated using a combination of in-situ techniques based on
the transverse momentum balance between a jet and a reference
object~\cite{Systematics_JES2}. The jets were required to have corrected
$\pT>25$~GeV and $|\eta|<$ 2.5.  Jets from pileup
were suppressed by requiring the absolute value of the jet vertex fraction\footnote{The
jet vertex fraction is defined using the tracks matched to a jet as the ratio
of the scalar sum of the transverse momenta of tracks from the main PV to that
of all tracks. A jet without any track matched is assigned a jet vertex fraction value of $-1$.} 
for jets with $\pT<50$~GeV and $|\eta|<2.4$ to be above 0.5.

Jets were $b$-tagged (identified as originating from $b$-quarks) using the
multivariate-based algorithm algorithm (MV1)~\cite{btag_2014}.  This is a neural network-based algorithm that
makes use of track impact parameters and reconstructed secondary vertices.
Jets were identified as $b$-jets by requiring the MV1 output to be above a
certain threshold value. This value was chosen such that the overall tagging
efficiency for $b$-jets originating from top-quark decays in MC
$\ttbar$ events is 70\%.  Tagging scale factors were applied to correct for
the difference in the tagging efficiency between MC simulation and data, including the
inefficiency scale factors applied when a jet was not tagged.
The scale factors were derived as
explained in Refs.~\cite{btag_2012-043,btag_2014,cltag_2014}.

To avoid double-counting objects in an event and to suppress leptons from
heavy-flavor decays, the following procedure to
remove overlaps was applied to the reconstructed objects: (1) removal of jets matched within
$\Delta R=0.2$ of electrons with $\ET>25$~GeV; (2) removal of electrons
matched within $\Delta R=0.4$ of jets with $\pT>25$~GeV; (3) rejection of
events where the selected electron shares an inner detector track with a
selected muon; and (4) removal of muons matched within $\Delta R=0.4$
of jets with $\pT>25$~GeV.  After these steps, the events were required to have
exactly one lepton (an electron or a muon) selected as above and 
matched to the trigger object that caused the event to be selected,
and at least three jets, of which at least
one was $b$-tagged. Events with a second lepton with $\pT>25$~GeV were
discarded.\footnote{
When the selection is applied to electrons as well as muons, the
thresholds refer to $\pT$ for muons and $\ET$ for electrons.}
Finally, the events had to exceed minimum values for the W transverse mass $m_{\mathrm{T}}(W)$ 
and the magnitude of the missing transverse momentum $\MET$.
The value of $\MET$ in the $\ttbar\to\ell\nu$+jets
process is relatively large due to neutrino(s) from $W$ decays.  It also
includes energy losses due to detector inefficiencies as well as
energy fluctuations in jet measurements.
The missing transverse momentum
value is calculated as the negative of the vector sum over the
energies of all clusters in the calorimeters, and $\MET$ is refined by 
the application of the object-level corrections for the contributions arising
from identified electrons, muons, and jets. The $W$-boson transverse mass
$m_{\mathrm{T}}(W)$ is defined as
$m_{\mathrm{T}}(W)=\sqrt{2\pT^{\ell}\pT^\nu(1-\cos(\varphi^{\ell}-\varphi^\nu))}$, where
$\pT^{\ell}$, $\varphi^{\ell}$ and $\pT^\nu$, $\varphi^\nu$ 
refer to the transverse component and $\varphi$ value of the lepton momentum and the missing momentum, respectively.
The events in this
analysis were selected by requiring $\MET>30$~GeV and
$m_{\mathrm{T}}(W)>30$~GeV.

After applying the selections listed above, the signal efficiency for
$\ttbar$ events with at least one top-quark decaying leptonically
was found to be 4.4\% in the $e$+jets channel and 5.5\% in the $\mu$+jets
channel.

While the majority of events in the selected sample include `real' 
leptons
(prompt electrons and muons from vector-boson decays), a small fraction of
events is due to leptons referred to as `fakes' 
(which include
nonprompt leptons and artifacts of the reconstruction).  There are several
sources of fake leptons.  Fake muons predominantly come from semileptonic
decays of heavy-flavor hadrons. In the $e$+jets channel, in addition to this
source, there are electrons originating from conversions as well as photons and
hadrons misidentified as electrons. The signal sample is composed of events
with either real or fake leptons:
\begin{equation*}
N=N_{\rm real}+N_{\rm fake}.
\end{equation*}

The estimation of $N_{\rm fake}$ from MC simulation is 
difficult due to large
uncertainties in the modeling of fake lepton production, and impractical due to
a tiny fake rate requiring 
huge simulated samples. Therefore, the
contribution of fakes was evaluated from data using the matrix
method~\cite{Aad:2010ey}.  In addition to events from the signal data sample
(labeled as ``tight'' events), a second (``loose'') set enriched with fake
leptons was defined by removing the lepton isolation requirement. The number of
events in each sample can be written as
\begin{equation*}
\begin{array}{l}
N^{\rm tight}=N^{\rm tight}_{\rm real}+N^{\rm tight}_{\rm fake},\\ [.5em]
N^{\rm loose}=N^{\rm loose}_{\rm real}+N^{\rm loose}_{\rm fake}.
\end{array}
\end{equation*}
If the fractions of loose events that are also tight events are known, the
number of events with fake leptons in the signal sample can be determined from
a linear system of two equations. Given the probabilities for real and fake
leptons that already passed the loose selection to also pass the tight
selection,
$\varepsilon_{\rm real}=N^{\rm tight}_{\rm real}/N^{\rm loose}_{\rm real}$ and
$\varepsilon_{\rm fake}=N^{\rm tight}_{\rm fake}/N^{\rm loose}_{\rm fake}$,
as well as the number of loose and tight events,
the number of tight events with a fake lepton is
\begin{equation*}
N^{\rm tight}_{\rm fake}=\frac{\varepsilon_{\rm real}N^{\rm loose}-N^{\rm tight}}{\varepsilon_{\rm real}-\varepsilon_{\rm fake}}\varepsilon_{\rm fake}.
\end{equation*}

The probability $\varepsilon_{\rm real}$ was determined with a tag-and-probe
method~\cite{Systematics_lepton1} based on the identification of a tight lepton and a loose lepton in
events originating from $Z\to ee/\mu\mu$ decays.  The fake lepton probability
$\varepsilon_{\rm fake}$ in the $\mu$+jets channel was derived for muons with
large track $d_0$ significance.\footnote{ $d_0$ is the distance of closest
approach  in the transverse plane
of a track to the primary vertex of the event; the $d_0$ significance
is the $d_0$ value divided by its fitted uncertainty.} In the $e$+jets channel,
$\varepsilon_{\rm fake}$ was determined in the $\MET<$ 20$\GeV$,
$m_{\mathrm{T}}(W)<$ 20$\GeV$ region, where the number of events was corrected
for the presence of real leptons. In both channels, the real and fake
probabilities were parameterized in terms of lepton $\pT$ and $\eta$, the leading
jet $\pT$, the distance between the lepton and the closest jet
($\Delta R_{\ell j}$), the ratio of the $\pT$ of the closest jet to
$\Delta R_{\ell j}$, the total number of jets, and the number of $b$-tagged jets.
The product of one-dimensional parameterizations was used, since there are too few events 
for a multidimensional parameterization. Based on this procedure,
each event in the loose set was assigned a weight

[$(\varepsilon_{\rm real}-1)\varepsilon_{\rm fake}/(\varepsilon_{\rm real}-\varepsilon_{\rm fake})$ for events with tight leptons and
$\varepsilon_{\rm real}\varepsilon_{\rm fake}/(\varepsilon_{\rm real}-\varepsilon_{\rm fake})$ for the rest],
and the weighted sample of
data events selected using the loose object criteria was taken as the multijet
contribution to the LHD background distribution.

The numbers of observed and expected events are shown in Table~\ref{tab:nev}.
Uncertainties on the numbers of expected $t\bar{t}$, single top, and diboson
events are derived from theoretical uncertainties on the respective production
cross-sections.  The uncertainty on $V$+jets events is evaluated based on a $\pm 4$\%
uncertainty on inclusive $V$+jets production and a $\pm 24$\% uncertainty on
each additional jet~\cite{berends}, giving a total uncertainty of $\pm 45$\%
for $V$+jets background. This large
uncertainty does not have an impact on the final result, since the normalization 
of the $V$+jets background is obtained from the data as described in Sec.~\ref{sec:analysis}.
The uncertainty on the multijet background
is evaluated by varying the parameters of the matrix method as described in
Sec.~\ref{sec:systematics}.
  
 \begin{table*}[hbpt]
\begin{center}
 \caption{\label{tab:nev}
 Observed numbers of events in the $e$+jets and $\mu$+jets channel
 together with estimated contributions from the $t\bar{t}$ signal
 and various background sources and associated uncertainties
 as described in the text.
 }
 \begin{tabular}{lcc}
 Event counts & $e$+jets                   & $\mu$+jets \\
 \hline
 $t\bar{t}$   &   123000$^{+6000}_{-7000}$   & 152000$^{+8000}_{-9000}$ \\
 $V$+jets     &  \ 38000$\pm$17000           & 47000$\pm$21000 \\
 Single top   &    12100$\pm$600             & 14900$\pm$800 \\
 Dibosons     &      710$\pm$40              & 880$\pm$50 \\
 Multijets    &     2800$\pm$2400            & 1500$^{+300}_{-1500}$ \\
 \hline
 Total        & 177000$^{+18000}_{-19000}$   & 216000$\pm$23000 \\
 \hline
 Data         & 176286                       & 220369 \\
 \end{tabular}
 \end{center}
 \end{table*}

\section{\label{sec:analysis}
Determination of the $t\bar{t}$ cross-section
}
The number of $\ttbar$ events in each channel and their combination was obtained
from the data using a template fit to the LHD distribution. The LHD function was
constructed using the projective likelihood method defined in Ref.~\cite{Hocker:2007ht}.
The LHD for event $i$ is defined as the ratio of the
signal $L_i^{\rm s}$ to the sum of signal and background likelihood functions
$L_i^{\rm s} + L_i^{\rm b}$:
\begin{equation*}
D_i = \frac{L_i^{\rm s}}{L_i^{\rm s} + L_i^{\rm b}},
\end{equation*}
where the likelihood functions $L^{\rm s}=\prod_j v_j^{\rm s}$, $L^{\rm b}=\prod_j v_j^{\rm b}$ are the
products of the probability density functions of kinematic variables $v_j$ for
signal and background, respectively. The variables used in the LHD were lepton
pseudorapidity $\eta_{\ell}$ and transformed aplanarity
\mbox{${\cal A}' = \exp(-8{\cal A})$}. The aplanarity ${\cal A}$ is defined
as 3/2 times the smallest eigenvalue of the momentum tensor
\mbox{$M_{ij}=\sum^{N_{\rm objects}}_{k=1} p_{ik}p_{jk}/\sum^{N_{\rm objects}}_{k=1} p^2_k$}, where $p_{ik}$ is the $i$-th
momentum component and $p_k$ is the modulus of the momentum of object $k$. The
objects included in the sum are leptons, jets, and $\MET$.
The choice of
variables followed the analysis in Ref.~\cite{ATLAStop7TeVPaper} and was
motivated by optimization in terms of best separation between the signal and
the background, the quality of the MC description of the variables in the data,
and reduced sensitivity to the jet-energy scale. As the fraction of signal
events in the selected data sample is large, two discriminating variables were
found to be sufficient to provide an adequate signal-to-background separation.

The $\eta_{\ell}$ and ${\cal A}'$ distributions for data and MC simulation are shown in
Fig.~\ref{fig:lhvars}.  The correlation between these variables is 
found to be negligible in both signal and background.  The LHD distributions for
signal and background events are compared with the data in Fig.~\ref{fig:lhd}.
The templates of LHD in the $e$+jets and $\mu$+jets channels are different due to the fact that
electrons in the barrel-endcap transition region are excluded and have
lower efficiency in the forward region, 
whereas the efficiency as a function of $\eta_{\mu}$ is quite smooth.

A weighted sum of the templates for the $\ttbar$ signal and all the backgrounds was
fitted to a binned LHD distribution in the data.  The contributions from single
top and diboson events were fixed according to their theoretical cross-sections.
The contribution from multijets events was obtained using
the matrix method.

The normalizations of the signal and $V$+jets (the dominant
background) were left as free parameters in the fit. In each channel $j$,
the fit was performed by minimizing
the negative log-likelihood function
for a Poisson model,
\begin{equation*}
\mathbb{L}_j = -{\ln L}_j = 2\sum_{i}\left(\nu_{ij}-n_{ij} + n_{ij}\ln n_{ij} - n_{ij}\ln\nu_{ij}\right),
\end{equation*}
summed over all bins $i$, where $n_{ij}$ is the 
observed number of events and $\nu_{ij}$ is
the expected number of events in each bin or channel.
The latter is defined as
\begin{equation*}
\nu_{ij} = p^{\ttbar} s_{ij} + p^V_j b_{ij} + q_{ij},
\end{equation*}
where $s_{ij}$, $b_{ij}$, and $q_{ij}$ are the predicted numbers of
events for the $\ttbar$ signal, $V$+jets, and small backgrounds
(single top, dibosons, and multijets), respectively, 
and $p^{\ttbar}$ 
and $p^V_j$ are the parameters of the fit.
The resulting value of the fit parameter $p^{\ttbar}$
was used to extract the number of $\ttbar$ events in the data.
Figure~\ref{fig:lhfit} shows the distribution of the LHD for data
and MC simulation. In both Figs.~\ref{fig:lhvars} and \ref{fig:lhfit},
the contributions from $t\bar{t}$ and $V$+jets production are normalized
according to the results of the LHD fit.

The channels were combined
by minimizing
$\mathbb{L}=\sum_j\mathbb{L}_j$
where the sum is taken over the $e$+jets and $\mu$+jets channels.

After the number of \ttbar\ events in data $N_{t\bar{t}}$ was obtained from
the fit, 
the \ttbar\ production cross-section in each channel was determined as
\begin{eqnarray}
\sigma_{\ttbar} &=& \frac{N_{t\bar{t}}}{ {\varepsilon}_{\ttbar} \times {\cal B} \times {\cal L}},
\label{eq:ttxs}
\end{eqnarray}
where ${\varepsilon}_{\ttbar}$ is the signal-reconstruction efficiency
(4.4\% in the $e$+jets channel and 5.5\% in the $\mu$+jets channel),
${\cal B}$ is the $\ttbar\to\ge 1$~lepton+jets branching ratio, and
${\cal L}$ is the total integrated luminosity.

\begin{figure}[hbpt]
\centering
\begin{tabular}{cc}
\includegraphics[width=0.375\linewidth]{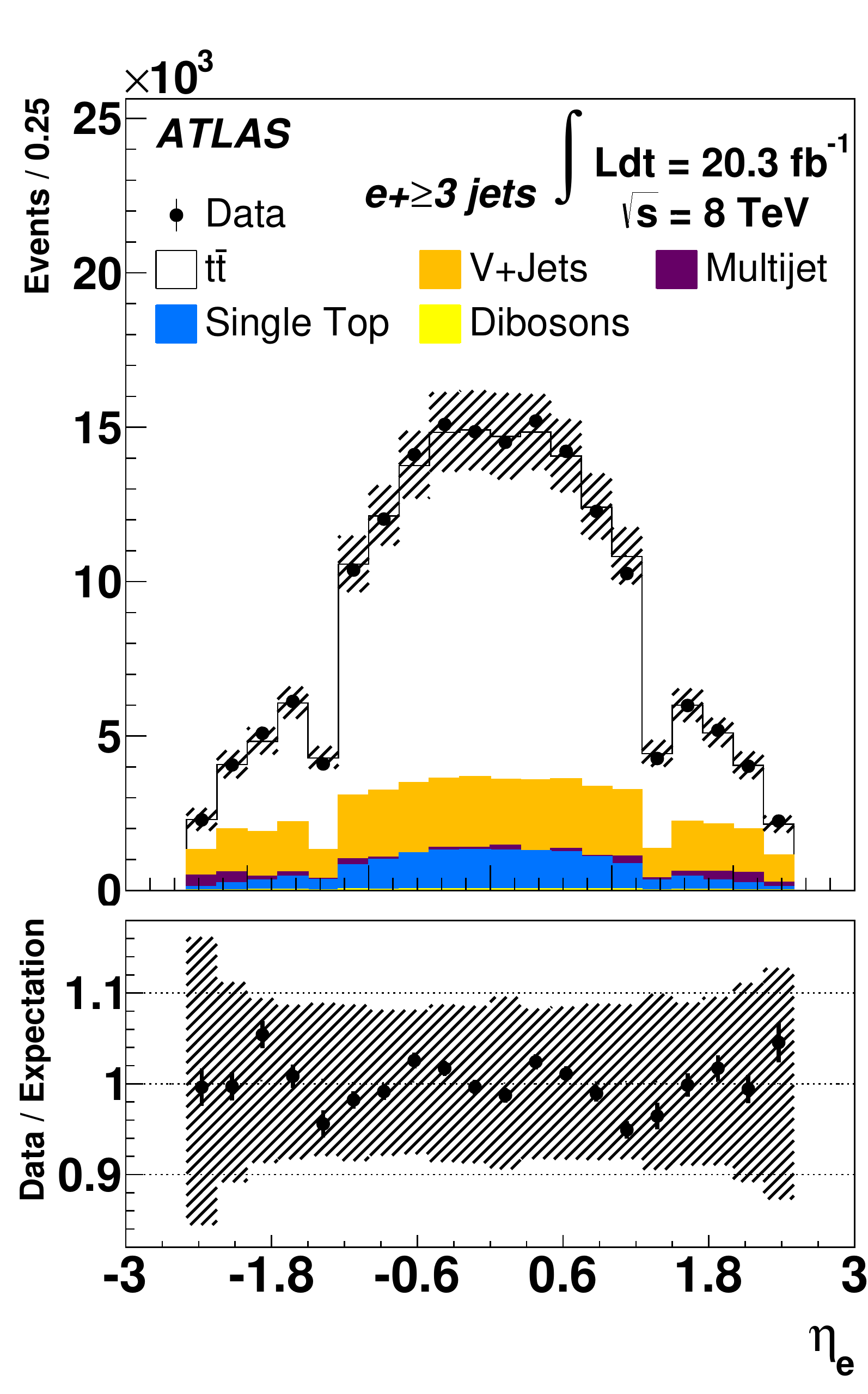}  &
\includegraphics[width=0.375\linewidth]{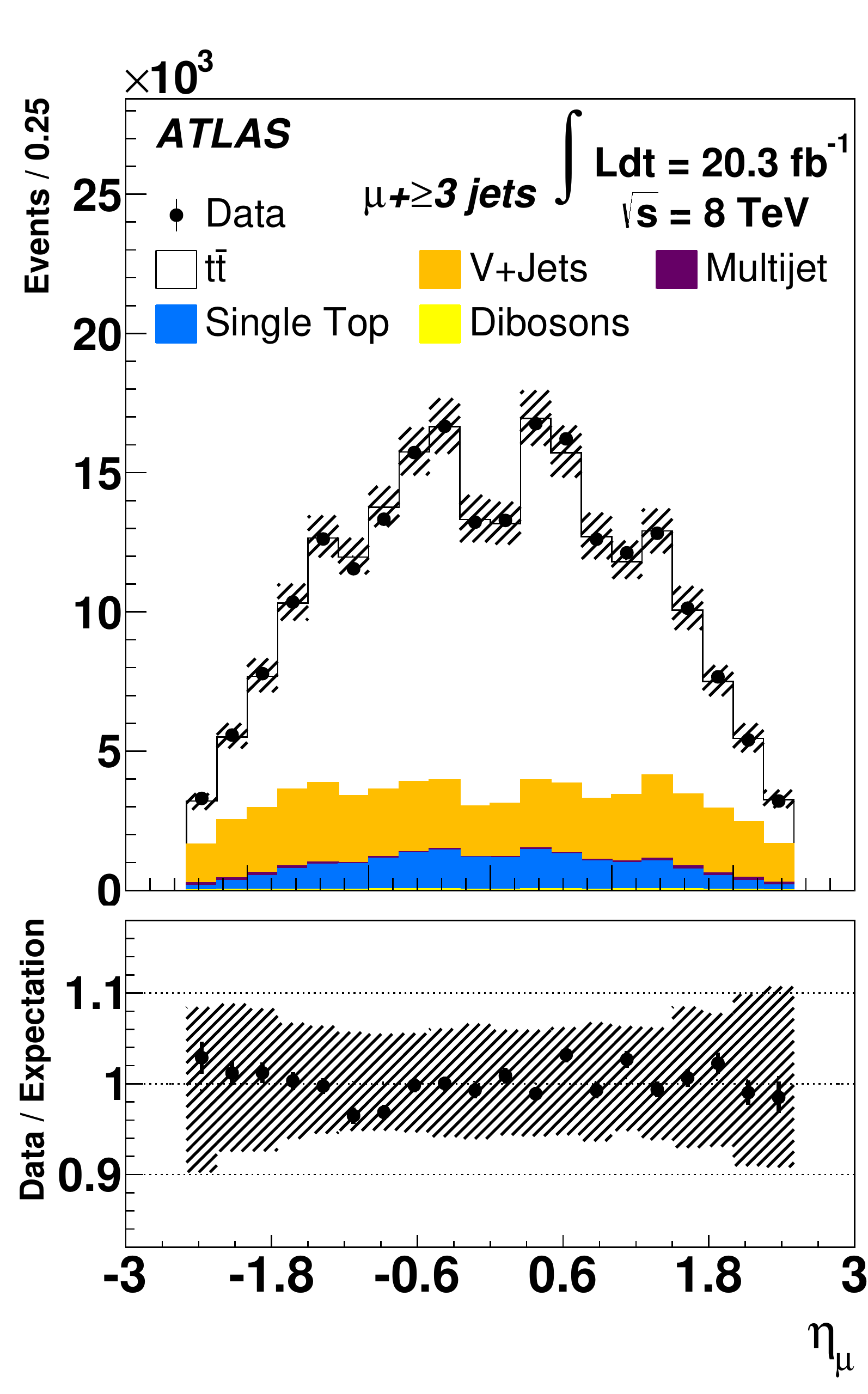}\\
\includegraphics[width=0.375\linewidth]{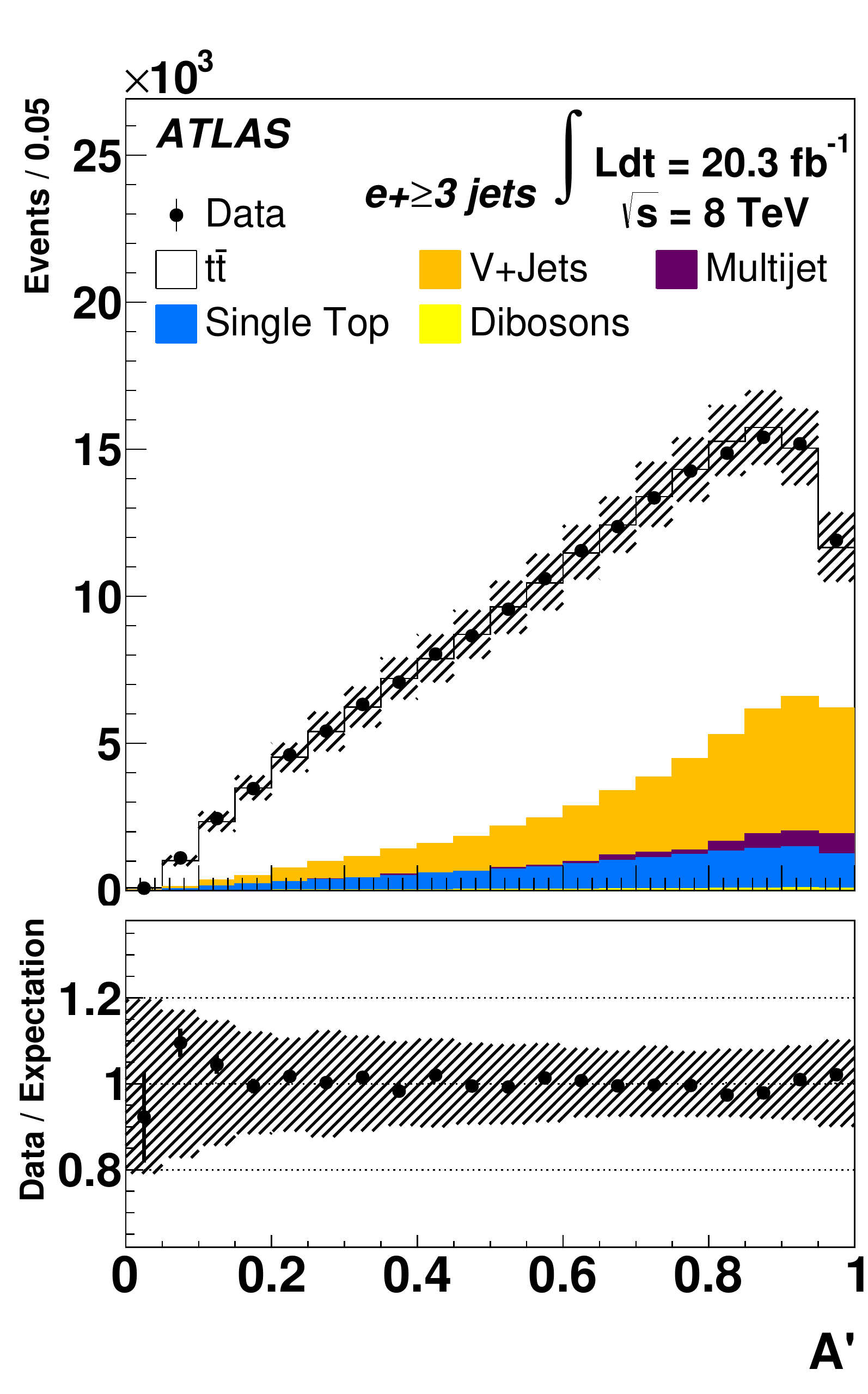}  &
\includegraphics[width=0.375\linewidth]{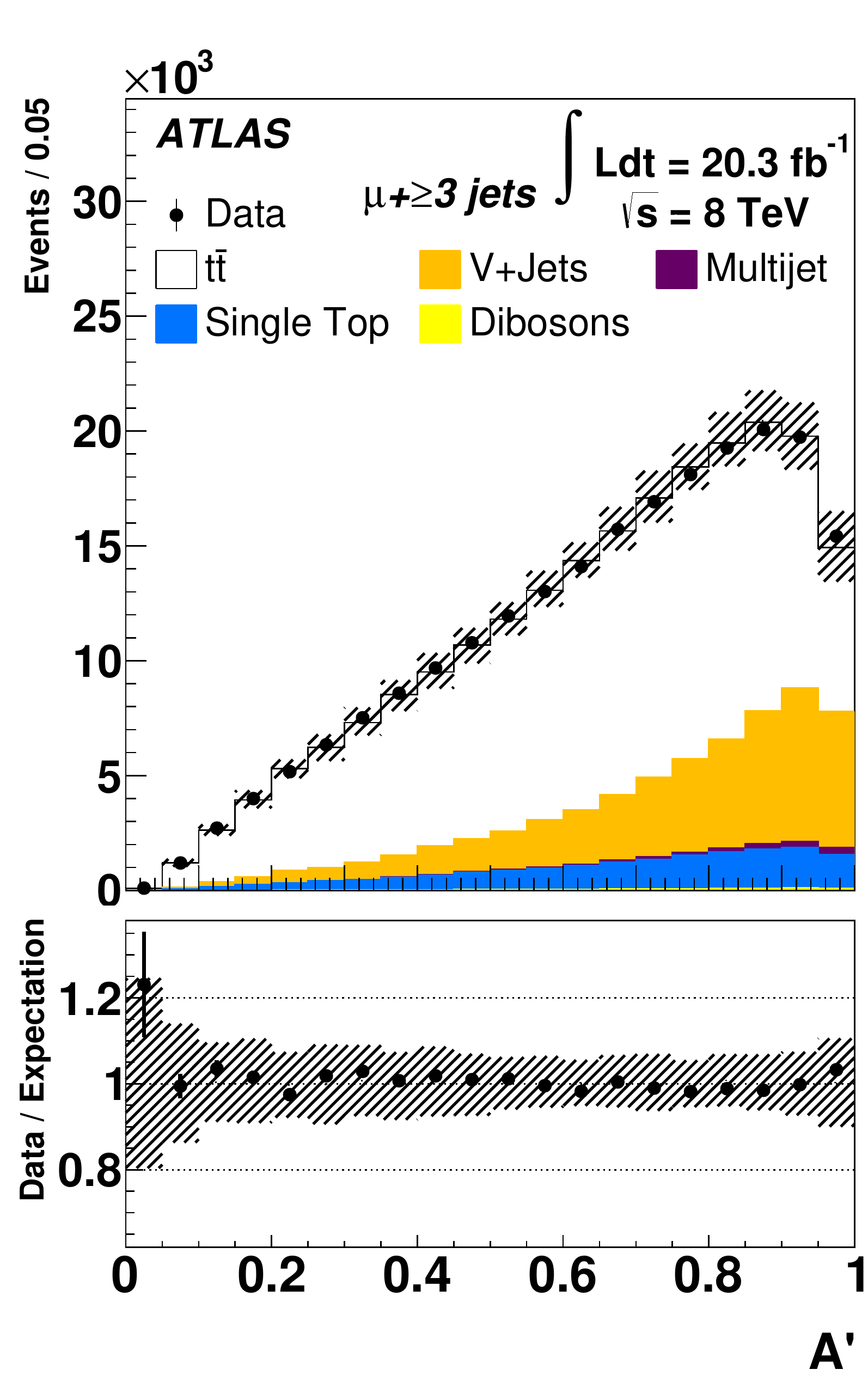} \\
\end{tabular}
\caption{\label{fig:lhvars}
Distributions of the lepton pseudorapidity (top) and
the modified aplanarity (bottom) in the $e$+jets (left) and
$\mu$+jets (right) channel. The data (dots) are
compared to the SM expectation broken down into contributions
from $t\bar{t}$, single top, $V$+jets, diboson, and multijet production.
The hatched areas correspond to
the combined statistical and systematic uncertainties excluding the
uncertainty on the $\ttbar$ signal modeling with different MC generators (see
Sec.~\ref{sec:systematics} for details).
The bottom part of each plot shows the ratio of the data
to the predicted value together with combined statistical and systematic
uncertainties.
}
\end{figure}

\begin{figure}[htp]
\centering
\begin{tabular}{cc}
\includegraphics[width=0.375\linewidth]{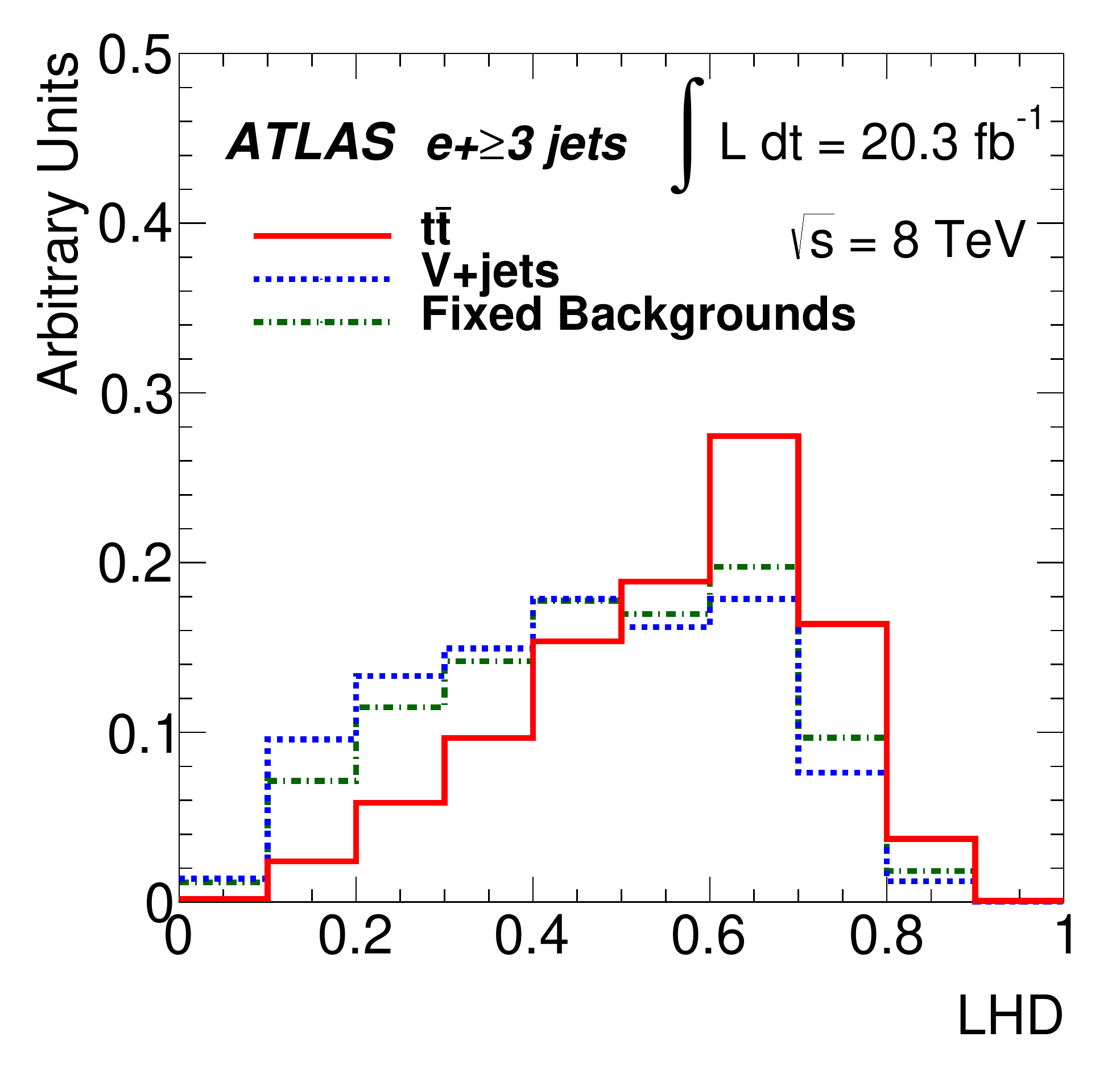}&
\includegraphics[width=0.375\linewidth]{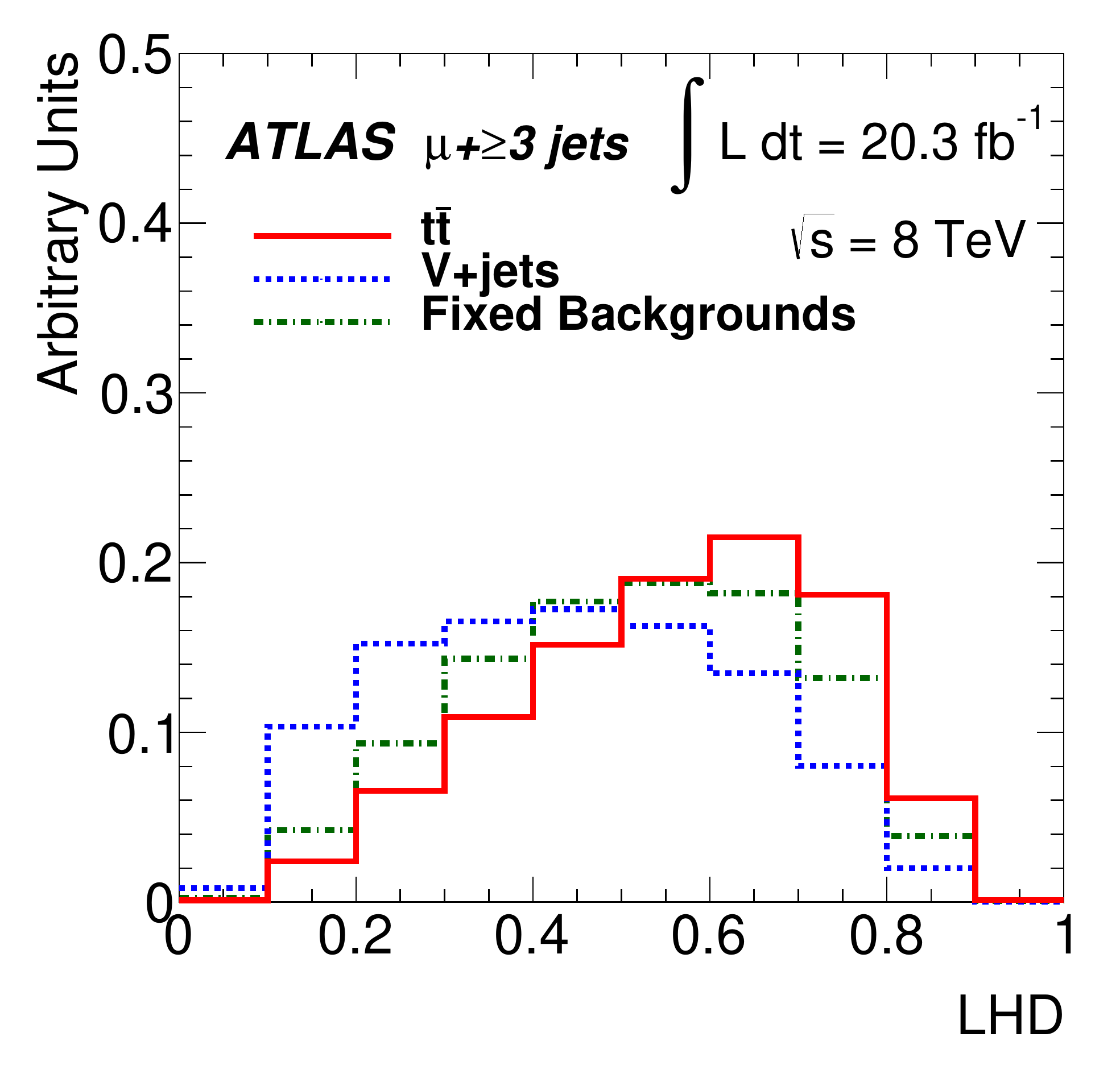}
\end{tabular}
\caption{\label{fig:lhd}
The templates of the likelihood discriminant LHD in the $e$+jets (left) and $\mu$+jets (right)
channels for signal, $V$+jets background, and fixed backgrounds, 
which include single top, diboson and multijets background.
}
\end{figure}

\begin{figure}[htp]
\centering
\begin{tabular}{cc}
\includegraphics[width=0.375\linewidth]{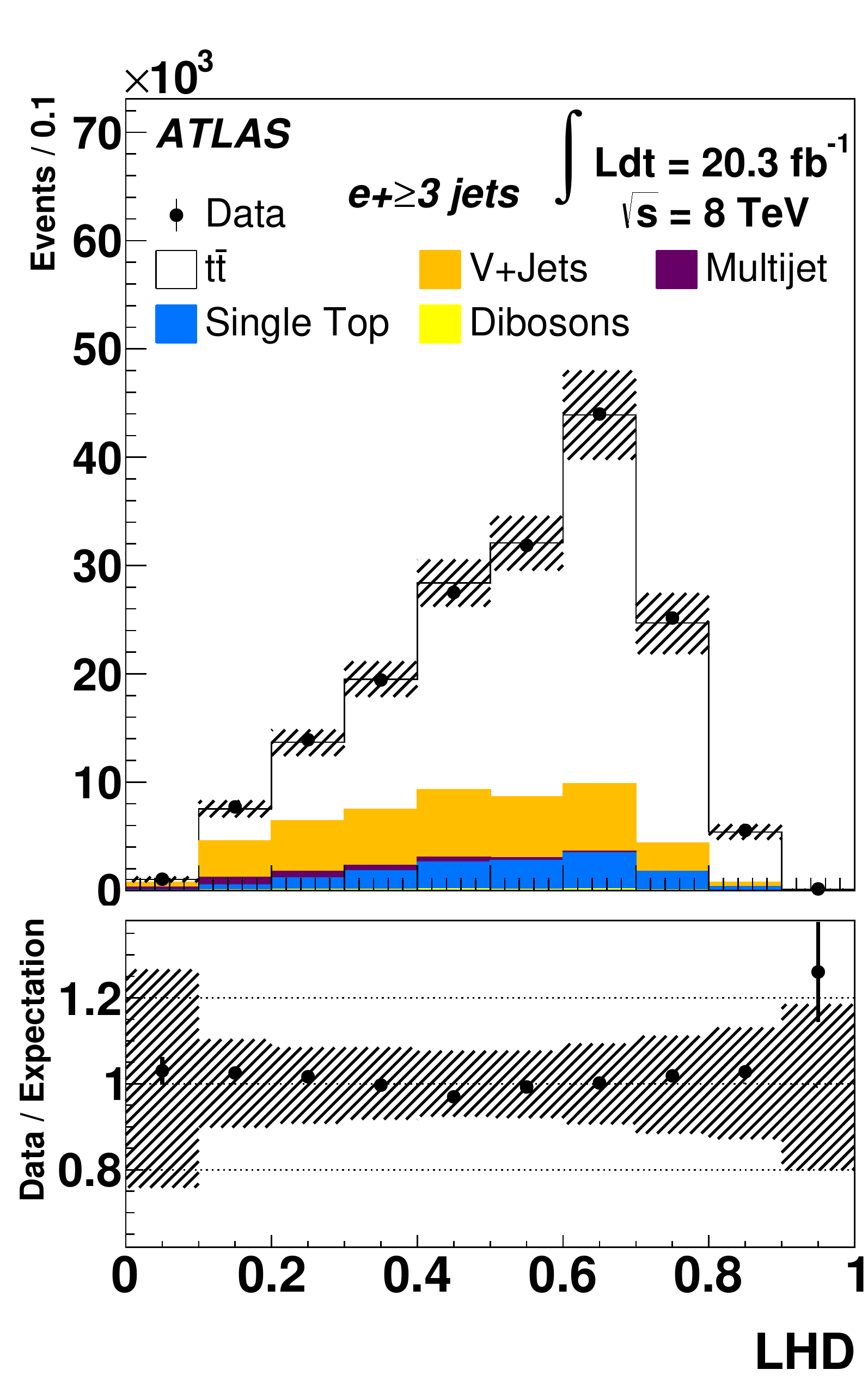}&
\includegraphics[width=0.375\linewidth]{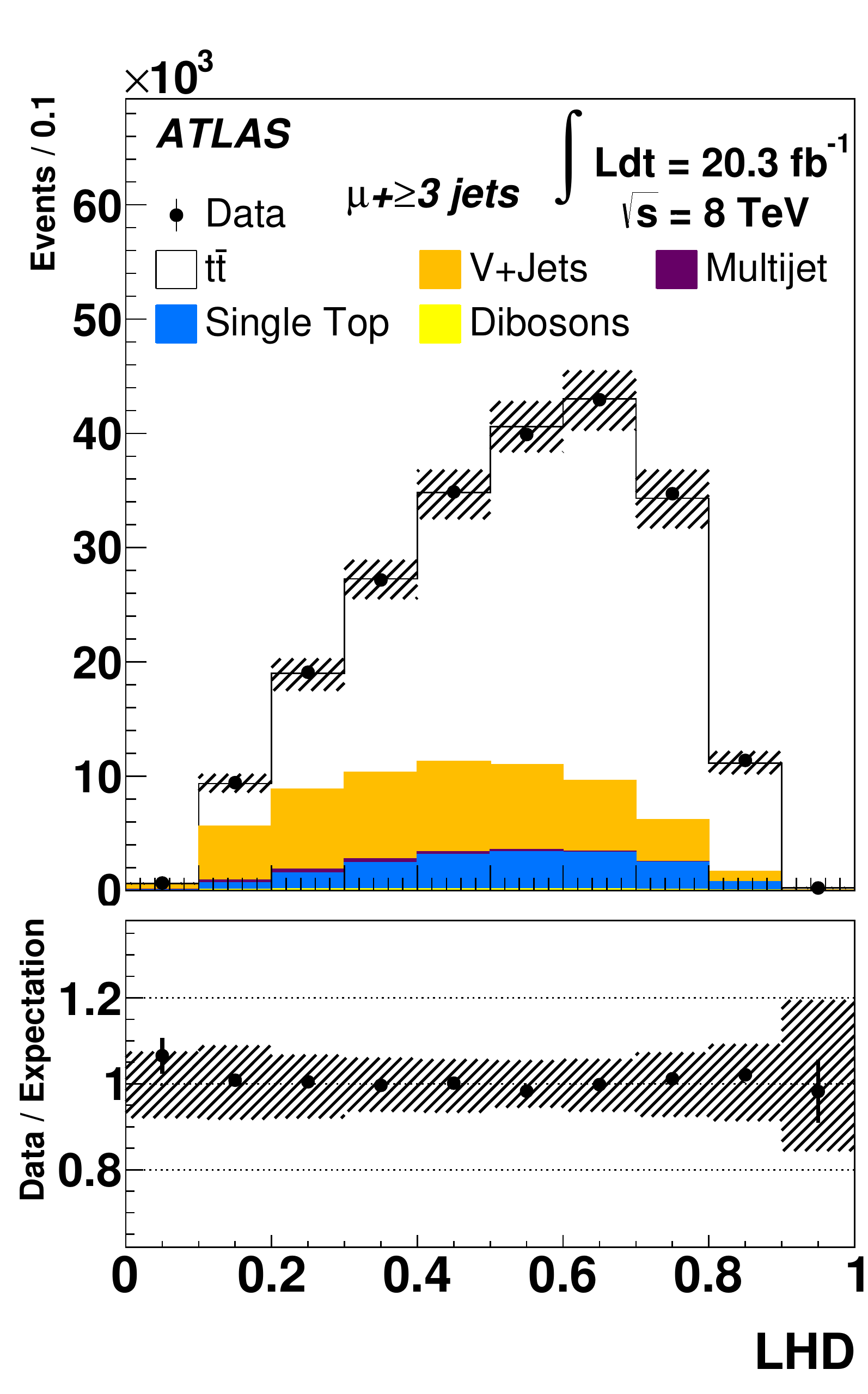}
\end{tabular}
\caption{\label{fig:lhfit}
Distributions 
of the likelihood discriminant LHD
in the $e$+jets (left) and $\mu$+jets (right)
data and a weighted sum of templates from the $t\bar{t}$
signal and various backgrounds.
The contributions from $t\bar{t}$ and $V$+jets production are normalized according to
the results of the likelihood discriminant fit, the single top and dibosons contributions are
normalized according to their theoretical cross-sections, and the
multijet background is normalized using the data-driven method explained
in the text.  
The bottom part of each plot shows the ratio of the data
to the predicted value together with combined statistical and systematic
uncertainties.
}
\end{figure}

\section{\label{sec:fiducial}
Fiducial cross-section measurement
}
The $\ttbar$ cross-section in the fiducial volume was measured to allow for a
more robust comparison to the theoretical prediction without extrapolating to
regions outside of the detector acceptance.  The definition of the
fiducial volume is based on MC simulation
and uses particle-level objects constructed using
stable particles with a mean lifetime $\tau_{\rm particle} > 0.3\times
10^{-10}$~s.  Electrons and muons were required to originate from
$t\to Wb\to \ell\nu\,b$ decays,
either directly or via a leptonically decaying $\tau$, and
to have $\pT>40$~GeV and $|\eta|<2.5$. The lepton momenta were corrected by
adding the energy and momentum of photons inside a cone of radius
$\Delta R=0.1$ around the lepton direction. Jets were created by
clustering particles using the anti-$k_t$ algorithm with
radius parameter $R=0.4$
(neutrinos, electrons and muons
from $W$-boson decays were excluded from particle-jet formation). Particles
from the underlying event were included in this definition, but particles
resulting from pileup were not. The particle-jets were required to have
$|\eta|<2.5$ and $\pT>25$~GeV.  The particle-jet $b$-identification was based
on the presence of nearby $B$-hadrons with a \pT of at least 5~GeV using the
ghost tagging method~\cite{Ghost_Asso_Jet}.  The overlap removal procedure
included the removal of particle-jets within $\Delta R=0.2$ of the nearest
electron with $\ET>25$~GeV followed by removal of electrons and muons  within
$\Delta R=0.4$ of the nearest remaining particle-jet with $\pT>25$~GeV.  The
events were required to have exactly one lepton with $\pT>40$~GeV, no second
lepton with $\pT>25$~GeV, and at least three particle-jets, of which at least
one jet was identified as a $b$-jet.  Missing transverse momentum was
calculated using neutrinos from $W$ leptonic decays. Similar 
to the reconstructed event selection, $\met>30$~GeV and $m_{\mathrm{T}}(W)>30$~GeV
requirements were applied.

The simulated $t\bar{t}$ events were split into two categories: in-fiducial
(satisfying the fiducial selection criteria) and out-of-fiducial (the rest).

The number of $\ttbar$ events $N_{t\bar{t}}$ obtained with the LHD fit was
converted to the number of $\ttbar$ events in the fiducial volume
$N_{t\bar{t}}^{\rm fid}$ using the fraction of reconstructed MC
events passing 
the fiducial selection. The number of $\ttbar$ events in
the fiducial volume was further used to determine the fiducial cross-section
$\sigma_{\ttbar}^{\rm fid}$ as
\begin{eqnarray}
\sigma_{\ttbar}^{\rm fid} &=& \frac{N_{t\bar{t}}^{\rm fid}}{ {\varepsilon}_{\ttbar}^{\rm fid}\times {\cal L}},
\label{eq:ttxs_fid}
\end{eqnarray}
where ${\varepsilon}_{\ttbar}^{\rm fid}$ is the fiducial signal-reconstruction
efficiency. The latter is defined as the ratio of the number of events passing
both the reconstruction requirements and the particle-level selection,
$N_{\rm reco}^{\rm fid}$, to the total number of events passing the particle
level selection, $N^{\rm fid}$:
\begin{eqnarray*}
{\varepsilon}_{\ttbar}^{\rm fid} &=& \frac{ N_{\rm reco}^{\rm fid} }{ N^{\rm fid}}.
\end{eqnarray*}
It was found to be 36\% in the $e$+jets channel and 44\% in the $\mu$+jets
channel.

\section{\label{sec:systematics}
Systematic uncertainties
}
The systematic uncertainties in the $\ttbar$ cross-section measurement can be
split into those related to the object reconstruction and momentum and energy
measurements, those related to background evaluation, and those related to the
signal modeling. The most important contribution to the total systematic
uncertainty arises from effects related to the $\ttbar$ modeling, as discussed
below.

To avoid statistical fluctuations, all systematic uncertainties were evaluated
using ensemble tests.  Simulated data sets (``ensembles'') were generated by
picking random combinations of events from a pool of MC original (nominal)
events according to the expected number of events due to signal and each
background type.  For each source of systematic uncertainty, the modified
events that passed the regular event selection were used to construct the LHD
templates used in the ensemble tests.  A weighted sum of modified signal and
background templates was fit to the LHD distributions for each ensemble.
The results of the fits were averaged over ensembles using a Gaussian
fit of the difference between the average number of
$t\bar{t}$ events for the nominal and the modified template fits, and
the fitted difference was propagated to the cross-section uncertainty.
The MC statistical uncertainty associated to the limited signal and
background sample size was found to be below $\pm$0.1\%.

{\bf Lepton reconstruction:} the uncertainties due to lepton trigger,
identification, energy or momentum resolution and reconstruction efficiencies
were estimated from $Z\to ee/\mu\mu$, $J/\psi\to ee/\mu\mu$, and $W\to e\nu$
processes using techniques discussed in
Refs.~\cite{Systematics_lepton0,Systematics_lepton1,Systematics_lepton2}.
These uncertainties are relatively small, dominated by the lepton
identification in the $e$+jets channel ($\pm$2\%) and the muon triggering efficiency
in the $\mu$+jets channel ($\pm$3\%). They are
specific to each lepton flavor and
therefore uncorrelated between the channels.

{\bf Jet reconstruction:} the uncertainty on the $\ttbar$ cross-section due to
the uncertainty on the jet-energy scale was estimated by varying the jet
energies according to the uncertainties derived from simulation and in-situ
calibration measurements using a model with
22 orthogonal
components~\cite{Systematics_JES,Systematics_JES2}.  The variations in jet
energies were also propagated to the $\MET$ value.  The uncertainty due to the
difference in jet-energy resolution between the data and MC events was evaluated by
smearing the MC jet transverse momentum according to the jet resolution as a
function of the jet $\pT$ and $\eta$~\cite{Systematics_JER}.  The uncertainty
due to the jet reconstruction efficiency was estimated by randomly discarding
jets according to the difference in jet reconstruction efficiency between the
data and MC.  The lower value of 
jet vertex fraction
was varied
between 0.4 and 0.6 as motivated by the $Z\to ee/\mu\mu$+jets
studies~\cite{JVF_2013-083}.  The dominant 
jet reconstruction systematic uncertainty on the
$\ttbar$ cross-section is due to the uncertainty on the jet-energy scale ($\pm$3\%).

{\bf $b$-tagging:}  the uncertainties on the $b$-tagging scale factors were
obtained separately for $b$-jets, $c$-jets and light flavor jets, and were treated as uncorrelated for 
each type of jet.
The uncertainties on the inefficiency scale factors,
applied when a jet was not tagged, were treated as fully anticorrelated to
those from the efficiency scale factors.  Since the efficiency to tag a $b$-jet
is 70\%, and only one of the two $b$-jets in the event is required to be
tagged, the $\ttbar$ event tagging efficiency is large (about 90\%) and its
uncertainty is dominated by $b$-jet tagging uncertainties, which are small
compared to non-$b$-jet ones. The resulting $b$-tagging contribution to the
overall systematic uncertainty is about $\pm$2\%.

{\bf Missing transverse momentum:} 
the systematic uncertainties associated with the momenta and energies of 
reconstructed objects (leptons and jets) were also propagated to the $\MET$ calculation.
The $\MET$ reconstruction also receives contributions from
the presence of low-$\pT$ jets and calorimeter cells not included in
reconstructed objects (``soft terms''). The systematic uncertainty on soft
terms was evaluated using $Z\to\mu\mu$ events from the $\MET$ data/MC ratio in
events without jets and from the balance between soft terms and hard
objects using methods similar to those used in Ref.~\cite{Aad:2012re}.
The $\MET$ measurement is not a significant source of systematic uncertainty
(below $\pm$1\%).

{\bf $W$+jets model:} this source of systematic uncertainty does not affect the
number of $\ttbar$ events expected after applying the selection requirements
but it does change the shape of the background template.  This uncertainty was
estimated using templates with $W$+jets events generated by {\sc Sherpa}
instead of {\sc Alpgen}+{\sc Pythia} and was found to be small (below 1\%) and
to partially cancel out when combining the $e$+jets and $\mu$+jets channels.

{\bf $W$ + heavy-flavor composition:} the heavy-flavor composition of $W$+jets
events was studied using $W$ events with two additional jets split into $W^+$
and $W^-$ samples.\footnote{The ratio of $W^+$ to $W^-$ production
cross-sections at the LHC is predicted much more precisely than the
cross-sections themselves~\cite{Kom:2010mv}.} Uncertainties were derived from
contributions from $W+c$, $W+c\bar{c}$, and $W+b\bar{b}$ based on the statistical
uncertainty, uncertainties on the extrapolation from 2-jets to $\ge$3-jets, and
MC modeling, following the procedure described in Ref.~\cite{Aad:2012hg}.  The
uncertainties themselves are very large (about $\pm$50\%), but the LHD template
shapes for $W$-bosons produced in association with different jet flavors are similar
and since the normalization of the $W$+jets background is derived from a fit to
the data, the resulting systematic uncertainty on the $\ttbar$ production
cross-section is small ($\pm$1\%).

{\bf Small backgrounds:}
uncertainties on single top-quark and diboson production were evaluated by
varying 
the relevant theoretical cross-sections within their uncertainties $\pm 1\sigma_{\rm theor}$,
which resulted in a $\ttbar$ production cross-section uncertainty of the order
of $\pm$1\%.  The uncertainty on the multijet contribution was derived from a
comparison with results with $\varepsilon_{\rm real}$ obtained using data
control samples with large $m_{\mathrm{T}}(W)$ ($\mu$+jets) or large $\MET$
($e$+jets) depleted of fake leptons; the variation of $\varepsilon_{\rm fake}$
due to the change of the low $m_{\mathrm{T}}({\rm W})$/$\MET$ region; and the
choice of MC samples used to subtract the real leptons. 
As seen in
Table~\ref{tab:nev}, the uncertainty on the number of multijet events is very
large (about $\pm$100\%), mainly due to uncertainties in $\varepsilon_{\rm fake}$,
but their fraction in the data sample is small due to the high lepton-$\pT$
requirement, so the overall effect is about $\pm$1\%.

{\bf Pileup:} the MC samples used in the analysis were
reweighted to reproduce correctly the distribution of the
number of $pp$ collisions per bunch crossing in the data. The accuracy of the
average correction factor due to pileup reweighting was found to be $\pm$4\%.
The pileup reweighting uncertainties originated from the modeling
of pileup events, including uncertainties in the $pp$ inelastic cross-section.
The resulting effect on the measured $\ttbar$ cross-section is small
(about $\pm$0.1\%) but is still included in the total systematic uncertainty.
 
{\bf Initial- and final-state radiation (ISR and FSR) modeling:} ISR or FSR changes the
number of jets in the event. To evaluate the uncertainty linked to 
the modeling of the ISR or FSR, $\ttbar$ MC samples
with modified ISR and FSR modeling were used.
The MC samples used for the
evaluation of this uncertainty were generated using the {\sc AcerMC} generator interfaced to {\sc Pythia},
where the parameters of the generation ($\Lambda_{\rm QCD}$, $Q^2_{\rm max}$
scale, transverse momentum scale for space-like parton-shower evolution~\cite{PYT6}) were varied to span the ranges compatible with
the results of 
a measurement of $\ttbar$ production with a veto on additional central jet activity~\cite{rapidity_gap}.
This uncertainty is large for the total inclusive $\ttbar$ production
cross-section ($\pm$3\%) but significantly reduced (below $\pm$1\%) for the fiducial
cross-section measurement.

{\bf MC generator:} the choice of MC generator used in the signal modeling
affects the kinematic properties of simulated $\ttbar$ events and reconstruction
efficiencies. For the purpose of addressing this effect,
$\ttbar$ events simulated with {\sc MC@NLO} interfaced 
to {\sc Herwig}+{\sc Jimmy} were used and compared to results obtained with {\sc Powheg}
interfaced to {\sc Herwig}+{\sc Jimmy}.
The resulting systematic uncertainty was found to be $\pm$3\%.

{\bf MC parton shower and fragmentation model:}
the effect of the parton shower and fragmentation modeling was estimated
by comparing the results obtained from the default MC sample simulated by
{\sc Powheg} interfaced to {\sc Pythia} with a modified model simulated by
{\sc Powheg} interfaced to {\sc Herwig}+{\sc Jimmy}. 
The change
in the number of
expected $\ttbar$ events due to the modified selection efficiency and the LHD
distribution was propagated to the systematic uncertainty on the $\ttbar$
production cross-section. Using this approach, the systematic uncertainty due
to the MC parton shower and fragmentation model was found to be $\pm$2\%.

{\bf MC parton distribution functions (PDFs):} this uncertainty is one of the
most significant ones in this analysis ($\pm$6\%).
An event-by-event reweighting was applied to the {\sc MC@NLO} $\ttbar$ sample
generated using the central value of the CT10 PDF. The CT10 variations, as well
as both the central values and the variations of MSTW2008 NLO at the 68\%
confidence level and NNPDF23 were used to evaluate the systematic uncertainty,
following the PDF4LHC recommendations~\cite{TTPDF2011}. For each PDF
and its variations the change in the $\ttbar$ cross-section was
calculated using ensemble tests. The final envelope of all variations was
symmetrized and half of its size was quoted as the systematic uncertainty due
to the PDF uncertainty. The effect arises primarily from the gluon PDF
component, which is large for signal and small for background. Therefore the PDF
uncertainty is dominated by the choice of PDF in the $\ttbar$ simulation. The
PDF uncertainty affects both the LHD shape and the selection efficiency, and is
significant for both the inclusive and the fiducial cross-section measurement.

{\bf Luminosity:} this uncertainty was evaluated separately and not combined
with the rest of the systematic uncertainties. Following the method described
in Ref.~\cite{lumi_7TeV}, 
the uncertainty on the total integrated luminosity was determined to be $\pm$2.8\% 
from a preliminary calibration of the luminosity scale derived from beam-separation scans performed in November 2012.
By combining the change in ${\cal L}$ in
Eqs.~(\ref{eq:ttxs}) and (\ref{eq:ttxs_fid}) and the change in the contribution of
single top and diboson events, the overall uncertainty due to the
integrated luminosity was found to be $\pm$2.9\%.

{\bf LHC beam energy:} calibrations determined the beam energy to be 
\mbox{($0.30 \pm 0.66$)}\% smaller than the nominal value of 4~TeV per
beam~\cite{LHC_beam_energy}.  The measured $\ttbar$ cross-section value is not
corrected for this shift. However, an uncertainty of $\pm$1.7\%, corresponding to 
the expected change in $\sigma_{\ttbar}$ for a $\pm$0.66\% change in $\sqrt{s}$ is
quoted separately in the final result. The effect of changing the LHC beam energy
on $\varepsilon_{\ttbar}$ is negligible.

The evaluation of the systematic uncertainties other than the $\ttbar$ MC
dependence for the measurement in the fiducial volume is the same as for the
inclusive measurement.  For the systematic uncertainties obtained using
different $\ttbar$ MC samples, 
the different signal-reconstruction efficiency in the fiducial volume 
was taken into account.

The systematic uncertainties due to
the jet reconstruction and $\met$ measurement are grouped together, as well as
the systematic uncertainties due to the background evaluation.
A summary of the systematic uncertainties is given in Table~\ref{tab:sys}.

\begin{table*}[hbpt]
\begin{center}
\caption{\label{tab:sys}
Summary of the systematic uncertainties in the measurements of the
$t\bar{t}$ production cross-section (\%). The bottom part of the table
lists systematic uncertainties for the fiducial $t\bar{t}$
production cross-section for the cases where
they are different from
those for the inclusive one.
}
\begin{tabular}{lccc}
Uncertainty on inclusive $\sigma_{t\bar{t}}$
& $e$+jets & $\mu$+jets & $\ell$+jets \\
\hline
Lepton reconstruction            &  $+2.7$~$-2.6$ &  $+2.1$~$-1.9$ &  $+1.7$~$-1.6$ \\
Jet reconstruction and $\met$      &  $+3.3$~$-3.9$ &  $+2.6$~$-3.2$ &  $+2.8$~$-3.4$ \\
$b$-tagging                      &  $+2.1$~$-1.9$ &  $+2.2$~$-1.9$ &  $+2.1$~$-1.9$ \\
Backgrounds                      &  $+2.8$~$-3.0$ &  $+1.8$~$-2.1$ &  $+1.7$~$-2.1$ \\
\hline
Monte Carlo generator
& $-2.2$~$+2.2$ & $-3.3$~$+3.3$ & $-2.7$~$+2.7$ \\ 
Parton shower and fragmentation
& $+2.0$~$-2.0$ & $+2.6$~$-2.6$ & $+2.3$~$-2.3$ \\ 
Initial- and final-state radiation
& $-4.1$~$+4.1$ & $-1.8$~$+1.8$ & $-3.0$~$+3.0$ \\ 
Parton distribution functions
& $+6.2$~$-6.0$ & $+5.6$~$-5.9$ & $+5.9$~$-5.9$ \\ 
\hline
Total
& $+9.7$~$-9.8$ & $+8.4$~$-8.7$ & $+8.6$~$-8.9$ \\ 
\hline
\hline
Uncertainty on fiducial $\sigma_{t\bar{t}}$
& $e$+jets & $\mu$+jets & $\ell$+jets \\
\hline
Monte Carlo generator
& $-2.1$~$+2.1$ & $-3.5$~$+3.5$ & $-2.8$~$-2.8$ \\
Parton shower and fragmentation
& $-2.6$~$+2.6$ & $-3.1$~$+3.1$ & $-2.9$~$+2.9$ \\ 
Initial- and final-state radiation
& $+0.4$~$-0.4$ & $+0.2$~$-0.2$ & $+0.3$~$-0.3$ \\ 
\hline
Total
& $+8.9$~$-9.0$ & $+8.5$~$-8.8$ & $+8.3$~$-8.6$ \\
\end{tabular} 
\end{center}
\end{table*}

\section{\label{sec:conclusions}
Results and conclusions
}
Inclusive and fiducial $\ttbar$ production cross-sections have been measured at
the LHC in $pp$ collisions at $\sqrt{s}$~=~8~TeV using data collected with the
ATLAS detector in 2012, corresponding to 20.3~fb$^{-1}$ of integrated
luminosity. The measurements were 
in the lepton+jets final state
using a likelihood discriminant fit.
Assuming a top-quark mass of
172.5~GeV, the inclusive $\ttbar$ production cross-section is found to be
\begin{align*}
e{\rm+jets}:
\sigma_{\ttbar}&=
256\pm 2({\rm stat.})~\pm 25({\rm syst.})~\pm 7({\rm lumi.})~\pm 4({\rm beam}){\rm~pb,}\\
\mu{\rm+jets}:
\sigma_{\ttbar}&=
260\pm 1({\rm stat.})~^{+22}_{-23}({\rm syst.})~\pm 8({\rm lumi.})~\pm 4({\rm beam}){\rm~pb,}\\
\ell{\rm+jets}:
\sigma_{\ttbar}&=
258\pm 1({\rm stat.})~^{+22}_{-23}({\rm syst.})~\pm 8({\rm lumi.})~\pm 4({\rm beam}){\rm~pb,}
\end{align*}
where the four quoted uncertainties are due to the number of events in data
(statistical uncertainty), systematic
effects (as described in Sec.~\ref{sec:systematics}),
the imprecisely known integrated luminosity, 
and the LHC beam energy.
Since the results were obtained using a fixed value of $m_{\rm top}$,
a set of $\ttbar$ samples with different top-quark masses was
generated in order to study the $m_{\rm top}$ dependence of the 
measured $\ttbar$ production cross-section, which is found to be
$(\Delta\sigma_{\ttbar}/\sigma_{\ttbar})/\Delta m_{\rm top}=-1.1$~\%/GeV
in all channels.

The fiducial $\ttbar$ cross-section is found to be
\begin{align*}
e{\rm+jets}:   \sigma_{\ttbar}^{\rm fid}&=11.3\pm 0.1({\rm stat.})~ \pm 1.0      ({\rm syst.})~\pm 0.3({\rm lumi.})~\pm 0.2({\rm beam}) {\rm~pb},\\
\mu{\rm+jets}: \sigma_{\ttbar}^{\rm fid}&=11.5\pm 0.1({\rm stat.})~ \pm 1.0      ({\rm syst.})~\pm 0.3({\rm lumi.})~\pm 0.2({\rm beam}) {\rm~pb},\\
\ell{\rm+jets}:\sigma_{\ttbar}^{\rm fid}&=22.8\pm 0.1({\rm stat.})~^{+1.9}_{-2.0}({\rm syst.})~\pm 0.7({\rm lumi.})~\pm 0.4({\rm beam}) {\rm~pb},
\end{align*}
where the four quoted uncertainties represent the same sources as
indicated before.

The largest systematic uncertainties are due to  $\ttbar$ MC modeling,
including the PDF uncertainty, the choice of MC generator, the parton shower and fragmentation model and
initial- and final-state radiation.  
The systematic uncertainty from ISR or FSR is found to be reduced for the fiducial cross-section 
measurement compared to the inclusive case, while the uncertainties due to the choice of MC generator and the parton shower and fragmentation 
model are found to increase slightly in the fiducial cross-section measurement compared to the inclusive case.

The measured inclusive $\ttbar$ production cross-section is
in good agreement with the NNLO+NNLL theory prediction, $\sigma_{t\bar{t}}=
253^{+13}_{-15}$~pb~\cite{TTcalc01}.
It is also consistent
with the ATLAS and CMS measurements of the $\ttbar$ production cross-section
in the dilepton channel~\cite{ATLAS_8TeV, CMS_8TeV}.
The ATLAS measurement had the result of 
$\sigma_{t\bar{t}} = 242.4 \pm 1.7({\rm stat.})~ \pm 5.5({\rm syst.})~ \pm 7.5 ({\rm lumi.})~\pm 4.({\rm beam})$~pb~\cite{ATLAS_8TeV},
where the four quoted uncertainties were due to the same sources as
above.

\section*{Acknowledgements}
We thank CERN for the very successful operation of the LHC, as well as the
support staff from our institutions without whom ATLAS could not be
operated efficiently.

We acknowledge the support of ANPCyT, Argentina; YerPhI, Armenia; ARC,
Australia; BMWFW and FWF, Austria; ANAS, Azerbaijan; SSTC, Belarus; CNPq and FAPESP,
Brazil; NSERC, NRC and CFI, Canada; CERN; CONICYT, Chile; CAS, MOST and NSFC,
China; COLCIENCIAS, Colombia; MSMT CR, MPO CR and VSC CR, Czech Republic;
DNRF, DNSRC and Lundbeck Foundation, Denmark; EPLANET, ERC and NSRF, European Union;
IN2P3-CNRS, CEA-DSM/IRFU, France; GNSF, Georgia; BMBF, DFG, HGF, MPG and AvH
Foundation, Germany; GSRT and NSRF, Greece; RGC, Hong Kong SAR, China; ISF, MINERVA, GIF, I-CORE and Benoziyo Center, Israel; INFN, Italy; MEXT and JSPS, Japan; CNRST, Morocco; FOM and NWO, Netherlands; BRF and RCN, Norway; MNiSW and NCN, Poland; GRICES and FCT, Portugal; MNE/IFA, Romania; MES of Russia and NRC KI, Russian Federation; JINR; MSTD,
Serbia; MSSR, Slovakia; ARRS and MIZ\v{S}, Slovenia; DST/NRF, South Africa;
MINECO, Spain; SRC and Wallenberg Foundation, Sweden; SER, SNSF and Cantons of
Bern and Geneva, Switzerland; NSC, Taiwan; TAEK, Turkey; STFC, the Royal
Society and Leverhulme Trust, United Kingdom; DOE and NSF, United States of
America.

The crucial computing support from all WLCG partners is acknowledged
gratefully, in particular from CERN and the ATLAS Tier-1 facilities at
TRIUMF (Canada), NDGF (Denmark, Norway, Sweden), CC-IN2P3 (France),
KIT/GridKA (Germany), INFN-CNAF (Italy), NL-T1 (Netherlands), PIC (Spain),
ASGC (Taiwan), RAL (UK) and BNL (USA) and in the Tier-2 facilities
worldwide.

\bibliographystyle{atlasBibStyleWoTitle}
\bibliography{a}
\clearpage
\newpage 
\begin{flushleft}
{\Large The ATLAS Collaboration}

\bigskip

G.~Aad$^{\rm 85}$,
B.~Abbott$^{\rm 113}$,
J.~Abdallah$^{\rm 152}$,
S.~Abdel~Khalek$^{\rm 117}$,
O.~Abdinov$^{\rm 11}$,
R.~Aben$^{\rm 107}$,
B.~Abi$^{\rm 114}$,
M.~Abolins$^{\rm 90}$,
O.S.~AbouZeid$^{\rm 159}$,
H.~Abramowicz$^{\rm 154}$
% ...
\bigskip
\\
% ...
$^{11}$ Institute of Physics, Azerbaijan Academy of Sciences, Baku, Azerbaijan\\
% ...
$^{85}$ CPPM, Aix-Marseille Universit{\'e} and CNRS/IN2P3, Marseille, France\\
% ...
$^{90}$ Department of Physics and Astronomy, Michigan State University, East Lansing MI, United States of America\\
% ...
$^{107}$ Nikhef National Institute for Subatomic Physics and University of Amsterdam, Amsterdam, Netherlands\\
% ...
$^{113}$ Homer L. Dodge Department of Physics and Astronomy, University of Oklahoma, Norman OK, United States of America\\
$^{114}$ Department of Physics, Oklahoma State University, Stillwater OK, United States of America\\
% ...
$^{117}$ LAL, Universit{\'e} Paris-Sud and CNRS/IN2P3, Orsay, France\\
% ...
$^{152}$ Institute of Physics, Academia Sinica, Taipei, Taiwan\\
% ...
$^{154}$ Raymond and Beverly Sackler School of Physics and Astronomy, Tel Aviv University, Tel Aviv, Israel\\
% ...
$^{159}$ Department of Physics, University of Toronto, Toronto ON, Canada\\
% ...
\end{flushleft}

\end{document}